\documentstyle[prl,aps]{revtex} \def\narrowtext{} \tighten \twocolumn
 
\begin{document}
\draft
 
\title{Evolution of the Fermi surface with carrier concentration
in Bi$_2$Sr$_2$CaCu$_2$O$_{8+\delta}$}
\author{
        H. Ding,$^{1,2}$
        M. R. Norman,$^2$
        T. Yokoya,$^3$
        T. Takeuchi,$^{2,4}$
        M. Randeria,$^5$
        J. C. Campuzano,$^{1,2}$
        T. Takahashi,$^3$
        T. Mochiku,$^{6}$ and K. Kadowaki$^{6,7}$
       }
\address{
         (1) Department of Physics, University of Illinois at Chicago,
             Chicago, IL 60607\\
         (2) Materials Sciences Division, Argonne National Laboratory,
             Argonne, IL 60439 \\
         (3) Department of Physics, Tohoku University, 980 Sendai, Japan\\
         (4) Department of Crystalline Materials Science, Nagoya 
             University, Nagoya 464-01, Japan\\
         (5)  Tata Institute of Fundamental Research, Bombay 400005, 
India\\
         (6) National Research Institute for Metals, Sengen, Tsukuba,
             Ibaraki 305, Japan\\
         (7) Institute of Materials Science, University of Tsukuba, 
             Ibaraki 305, Japan\\
         }

\address{%
\begin{minipage}[t]{6.0in}
\begin{abstract}
We show, by use of angle-resolved photoemission spectroscopy, that 
underdoped
Bi$_2$Sr$_2$CaCu$_2$O$_{8+\delta}$ 
appears to have a large Fermi surface 
centered at $(\pi, \pi)$, even for samples with a
$T_c$ as low as 15 K. 
No  clear evidence of a Fermi 
surface pocket around $(\pi/2, \pi/2)$ has been found.
These conclusions are based on a determination of the
minimum gap locus in the pseudogap regime $T_c < T < T^*$,
which is found to coincide with the 
locus of gapless excitations in momentum space (Fermi surface)
determined above $T^*$.
These results suggest that the pseudogap is more
likely of precursor pairing rather than magnetic origin.
\typeout{polish abstract}
\end{abstract}
\end{minipage}}

\maketitle
\narrowtext

One of the most significant facts that has been learned about high 
temperature
superconductors with optimal doping (highest $T_c$) is that they possess a
large Luttinger Fermi surface containing $1+x$ holes, where $x$ is the 
number of doped holes, rather than $x$.
This has been clearly established now for a number of cuprates 
by use of angle-resolved photoemission spectroscopy (ARPES),
including YBCO\cite{YBCO}, BSCCO\cite{BSCCO,DING96}, and NCCO\cite{NCCO}.  
Fewer studies exist on underdoped cuprates\cite{LIU}.
Recently, the Stanford group\cite{SHEN} and our
own group\cite{NATURE} have presented data on underdoped BSCCO.  
These data are quite remarkable in that they show the existence of a highly 
anisotropic (d-wave-like) pseudogap which
persists up to a temperature $T^*$ that can be significantly 
higher than $T_c$\cite{NATURE}.  

In this paper, we address in detail the question of the evolution
of the Fermi surface with doping. In lightly underdoped
samples, where $T^*$ is low
enough, we determine the Fermi surface --
the locus of points in momentum space with gapless excitations --
above $T^*$ where the pseudogap has vanished.
We find that this coincides with the locus of minimum excitation gap 
in the pseudogap regime ($T_c < T < T^*$). 
In the highly underdoped low $T_c$ samples,
$T^*$ is too high, so we have to infer
a Fermi surface from the minimum gap locus in the pseudogap regime.   
For all doping values studied, we find a large Fermi surface
whose volume is consistent with $1+x$ holes.  

The experiments were carried out at the 
Synchrotron Radiation Center
following procedures described 
previously\cite{NATURE,DING95}.  Underdoping 
was achieved by adjusting the oxygen partial pressure during annealing 
the float-zone grown crystals. These crystals have sharp x-ray diffraction 
rocking curves with structural coherence lengths
similar to the near-optimally doped samples studied earlier.
All samples show a very flat surface after cleaving, 
which is essential for determining the Fermi surface.
We have studied several samples ranging from overdoped to underdoped
(optimal doping corresponds to a 92K $T_c$);
in this paper we will focus on a moderately underdoped sample with
a 83K $T_c$ (transition width 2 K), 
and a heavily underdoped sample with $T_c$ of 15K
(width $>$ 5K), and
contrast these results with a slightly overdoped 87K sample (width 1K).
We will label the samples by their onset $T_c$'s. 

The first point to address is how to determine the Fermi surface.
Typically in ARPES,
one measures energy distribution curves (EDCs) at several k points along
various directions in momentum space.
Along a cut moving from occupied
to unoccupied states, a broad spectral peak in the EDC at high binding 
energy moves to lower binding energy and narrows as
the Fermi crossing is approached.  For $k > k_F$, the
spectral weight of the peak then drops off.  The Fermi crossing is usually
characterized by where the intensity of the peak has been reduced by half.
A more proper definition \cite{NK} is to realize that if a spectral function
interpretation of the EDC is correct, then by integrating the EDC over
energy, one obtains the momentum distribution function, $n_{\vec k}$.
$k_F$ is then defined to be where $|\nabla n_{\vec k}|$
is maximal.
We have shown, at least along certain cuts in
momentum space, that this procedure works for both YBCO
and BSCCO\cite{NK,PHMIX}.  
However, we have found that background emission,
superlattice effects, and $\vec k$-dependent
matrix elements can complicate
the identification of integrated EDCs with $n_{\vec k}$.

A simpler way to characterize the Fermi surface is to track the evolution of
the leading edge of the EDC.  If all EDCs along a cut
are plotted on top
of one another (see Fig.~1a), then as one moves
from occupied to unoccupied states, the leading edge moves 
to low binding energy, then moves back to higher binding energy.  A similar
behavior is found in simulations which take into account the
experimental dispersion and resolution, with the moving back of the
leading edge due to a drop of spectral intensity as $k$ goes beyond $k_F$.
The simulations find that the EDC just before the leading edge moves
back agrees with the actual Fermi crossing to within half the $k$-resolution.
The Fermi crossing determined this way 
typically coincides with the point where the leading edge 
has the sharpest slope and whose midpoint has moved to the lowest binding
energy.  
This criterion has the advantage of being well-defined
even when an excitation gap is present, in which case it defines 
the minimum gap locus.  We note that in a superconductor
the minimum gap locus coincides with the normal state Fermi
surface\cite{SCHRIEFFER,PHMIX}.  This would not be the case if the gap 
were due to a charge or spin density wave, since in
that case the minimum gap locus is determined by the q vector of the
instability, which would only coincide with the Fermi surface if one had
perfect nesting.

In Fig.~1a we show EDCs from two 
representative cuts in the
zone for the 83K sample at $T= 160$K.
Since the pseudogap closes at $T^* \simeq 170$K
for this sample\cite{NATURE}, these data essentially represent the
gapless state.  We note that the spectra are very broad, a
consequence of self-energy effects
further enhanced by the elevated
temperature. In effect, there are
no well-defined quasiparticles at this temperature.  Despite this, we
clearly see dispersion and infer Fermi crossings, where for $k > k_F$,
the spectral intensity decreases and the leading edge pulls back. 
The solid triangles plotted in Fig.~1c represent our
estimate of the Fermi surface which
is similar to that 
for optimally doped samples\cite{DING95,DING96,GAP}.  Most 
importantly, we infer a Fermi crossing along the $\bar{M}-Y$ ($\pi,0-\pi,\pi$)
direction.

Several cuts normal to the Fermi surface
have been measured in the pseudogap regime
at 90K, just above $T_c$ but well below $T^*$.
In the left panel of Fig.~1b we plot EDCs for one of
these cuts, and
show that the leading edge evolution is similar to that of Fig.~1a, the
difference being the presence of an excitation gap at 90 K.  
In the right panel,
we plot the midpoint shift of the leading edge of the spectrum and its width
along this cut
(determined by fitting the leading edge to
a Fermi function, the width
being the effective temperature of the fit).
Both the midpoint shift and the width of the leading edge have a 
minimum value at the same location along a cut, which defines the minimum
gap locus.  As shown in Fig.~1c, this
locus coincides with the 
Fermi surface above $T^*$ within experimental error bars.

In Fig.~2a we show EDCs from one of seven cuts (labeled I in Fig.~2c)
for the 15K sample at $T$=15 K, which shows strong dispersive
effects even in this highly underdoped sample.  The resulting
minimum gap locus from these cuts is shown in Fig.~2c.  This
locus is similar
to that of the 83K sample shown in Fig.~1c, and we find no evidence
for a small hole pocket centered about $(\pi/2,\pi/2)$.
To further test this,
we took a cut (labeled II in Fig.~2c) along $\bar{M}-Y/2$ ($\pi,0-\pi/2,\pi/2$),
whose EDCs are shown in Fig.~2b. 
If a pocket existed, then one would observe
a gapless point along this cut.  However a large gap is
found at all points, inconsistent with the existence of a Fermi crossing.
A plot of the leading edge midpoints along the minimum gap locus is 
similar to that of a 10K sample shown in our earlier 
work\cite{NATURE}, indicating the presence
of a highly anisotropic gap.  The gap monotonically increases
from the $(\pi,\pi)$ direction, but the behavior around the node is flattened
relative to a simple d-wave gap.  This flattening 
might suggest that the highly underdoped materials actually possess a 
gapless Fermi arc\cite{SHEN}.
However, as explained above, there is no evidence for the closing
of this arc, or for a shadow-band feature in the dispersion. 
A more likely explanation is that the increased linewidth
broadening in the underdoped material is responsible for this 
flattening near the node
(e.g., for Lorentzian broadening, the midpoint shift is equal to
the actual gap minus $\Gamma$), and that this small gap region is just
part of a large minimum gap locus which intersects $\bar{M}-Y$.  
We note that it becomes increasingly
difficult to determine the actual 
minimum gap locus in a region near the
$\bar{M}$ point as the sample becomes increasingly underdoped.  This is because
of the small dispersion near this point, which occurs even for optimal doping,
and the strong drop of the signal relative to background emission
as the doping decreases.
However, our data indicate that the leading edge
evolution along cuts near $\bar{M}$ in the underdoped samples is similar to
that at optimal doping and
consistent with the existence of a large minimum gap locus, which
we identify with the Fermi surface underlying the pseudogap state.

In Fig.~3 we plot the Fermi surfaces for the 87K, 83K, and 15K samples.
We emphasize that a large surface is found in all three cases, similar
to results obtained for underdoped YBCO\cite{LIU}.
Also plotted in Fig.~3 are the predicted surfaces based on a 
rigid band shift relative to the 87K sample for doping values of 
18\% (87K), 13\% (83K), and 6\% (15K). It is apparent from Fig.~3 that
the experimental error bars do not permit us to 
determine the changes in area enclosed by the Fermi surface,
and thus to deduce the hole doping $x$.
The error bars are due to determining
the Fermi crossings as well as the alignment of the samples.

We now turn to the issue of sample aging.
BSCCO has a natural
cleavage plane due to weak van der Waals coupling of the two BiO layers.
The BiO surface in optimally doped samples is relatively inert to
the residual gases in the vacuum chamber ($5\cdot10^{-11}$ Torr), and
a clean surface can be
regenerated by warming the sample above 50 K.  However, we have 
found
this not to be the case in our underdoped samples.  In particular, after a
day in the chamber, the quasiparticle peak at low temperatures becomes
narrower and stronger as illustrated in Fig.~4a, where we show the
spectrum of the 83K sample measured at $\bar{M}$ for a newly 
cleaved
surface (blue line) and after 41 hours (red line), with an 87K spectrum
(black line) plotted for comparison.
This is in contrast with expected behavior of surface 
contamination which would lead to a broader peak due to scattering of the
photoelectrons off adsorbed gases.  In fact, the
spectrum of the aged 
83K sample becomes similar to that of the 87K sample.  This leads us to 
speculate that the underdoped BSCCO surface's doping level has increased.
Another important finding from Fig.~4a is that 
the whole spectrum shifts toward the Fermi energy.  The
shift in the main valence band below 1 eV is about 40 meV which suggests a
rigid band shift due to an increase in the doping level.
Also, the aging behavior of underdoped BSCCO supports the claim that the 
broadness of quasiparticle peaks is due to underdoping rather than 
impurities\cite{NATURE}.  Similar behavior is seen in the 15K sample,
where the aged spectra show a sharper leading edge.  In fact, the leading
edge behavior of the aged 15K sample clearly shows that the Fermi surface
expands with aging in support of an increase in the doping level, as plotted
in Fig.~2c.
All data presented in Figs.~1-3 are 
obtained within a short period of time after cleaving, ensuring
that intrinsic spectra were measured.

Finally, we consider the theoretical implications of this work.
The coincidence of the Fermi surface with the
minimum gap locus is in support of a precursor pairing gap
interpretation of the pseudogap\cite{RANDERIA,FUKUYAMA},
but the present data cannot distinguish between pairing
of neutral spinons versus real electrons.
We should remark that
a simple mean-field simulation of a magnetic gap has a minimum gap 
location
along $\bar{M}-Y$, which does coincide with the paramagnetic Fermi crossing.  On
the other hand, this gap is around 200 meV if one requires a small hole
pocket with the desired size (x=0.06).  Moreover, the occupied band along
this direction is a shadow band (the main band being pushed above the Fermi
energy).  This is inconsistent with the fact that no
evidence of the shadow band is seen along $\Gamma-Y$.
We should remark that Marshall et
al\cite{SHEN} claim the existence of an energy scale at about
200 meV near $\bar{M}$ in underdoped samples.  In our opinion, this is
simply associated with the incoherent part of the spectral function.
A similar effect is seen in optimally 
doped
samples at low T due to the low frequency suppression of the imaginary part
of the self-energy caused by the superconducting gap, which produces a 
higher
binding energy hump in the spectrum separate from the quasiparticle
peak\cite{NK}.  In the normal state, this energy scale is
no longer apparent since the gapping effect is removed, and the
spectral peak location in the normal state is consistent with a small 
binding energy (Fig.~2, Ref.~\onlinecite{NK}).  To illustrate this further,
we show in Fig.~4b data from an underdoped 60K sample near $\bar{M}$.
The unaged EDC shows a
feature near 200 meV, but also clearly shows, at a much lower binding energy,
a sharp leading edge with a midpoint of 20 meV.  As the sample ages, the
200 meV feature moves up in binding energy (becoming the hump in optimally
doped samples), and the sharp leading edge evolves into a quasiparticle
peak.\cite{QP}

In conclusion, we find a large Fermi surface for underdoped BSCCO samples,
even with a $T_c$ as low as 15 K.  No evidence is found for small hole
pockets.  Moreover, we have shown, at least in an 83K sample,
that the minimum gap locus in the pseudogap phase coincides with the Fermi
surface determined in the normal state above $T^*$.  This
supports the idea that the pseudogap is of precursor pairing
rather than magnetic origin.

This work was supported by the U. S. Dept. of Energy,
Basic Energy Sciences, under contract W-31-109-ENG-38, the National 
Science Foundation DMR 9624048, and
DMR 91-20000 through the Science and Technology Center for
Superconductivity, the Japan Society for the promotion of Science, NEDO,
and the Ministry of Education, Science and Culture of Japan.
The Synchrotron Radiation Center is supported by NSF grant DMR-9212658.

\begin{figure}
\caption{Measurements of the 83K sample.
Fig.~1a: EDCs at a photon energy of 19 eV
at 160K along cuts I, II in Fig.~1c. 
Fig.~1b:  EDCs at a photon energy of 22 eV
at 90K along cut III in Fig.~1c (left panel).
Midpoint shifts (blue dots) and widths (red diamonds)
of this cut (right panel).
Fig.~1c:  Fermi surface (FS) at
160K (solid blue triangles) and
minimum gap locus (MGL) at 90K (solid red dots). Cuts I, II 
(open blue triangles) and
cut III (open red dots) are locations 
of EDCs in Fig.~1a and b.  
Notice that the two surfaces coincide within error bars.
The error bars represent uncertainties of Fermi crossings as well 
as possible sample misalignment.  The red curve is a rigid
band estimate of the Fermi surface.
}
\label{fig1}
\end{figure}

\begin{figure}
\caption{Measurements of the 15K sample at a photon 
energy of 22 eV at 15K.
Fig.~2a: EDCs along cut I in Fig.~2c. The red curve is the Fermi crossing,
and the blue ones umklapps due to the superlattice\protect\cite{DING96}.
Fig.~2b: EDCs along cut II in Fig.~2c. Note that a gap is present for
all EDCs from this cut.
Fig.~2c: The minimum gap locus (solid red dots).  Solid blue triangles represent
the minimum gap locus after the sample has aged, indicating an
increased doping (for visual clarity, it is shown only near the
$(\pi,\pi)$ direction).  Open dots and triangles
represent the cuts from Fig.~2a and b (with solid blue
dots marking the umklapp locations).
}
\label{fig2}
\end{figure}

\begin{figure}
\caption{Fermi surfaces of the 87K, 83K, and 15K samples (error bars
are similar to Figs.~1c and 2c). 
All surfaces have a large volume.
The solid lines are tight binding estimates of the Fermi surface
at 18\%, 13\%, and 6\% doping assuming rigid band behavior.
}
\label{fig3}
\end{figure}

\begin{figure}
\caption{Aging effects (T=14K).  
Fig.~4a: EDCs of the 83K sample at $\bar{M}$ for a newly 
cleaved surface (blue line) and after 41 hours (red line), with the EDC
of the 87K sample (black line) plotted for comparison. 
The 40 meV shift is between the leading edges of the main valence bands 
of the two 83K spectra.
Fig.~4b: EDC of the 60K sample near $\bar{M}$
for an unaged surface (blue line) and an aged surface (red
line).  Note the 200 meV hump moves to lower binding energy, and the low
binding energy leading edge evolves into a quasiparticle peak.
}
\label{fig4}
\end{figure}


\begin{references}
\bibitem{QP}
We do not see a sharp quasiparticle peak in the unaged spectrum of the
60K sample, even at 14K.  This may be due to the very broad transition
width of this sample.
\end{references}
\end{document}